\documentclass{osa-article}

\journal{oe}
\usepackage{float}

\begin{document}

\title{Single-channel electronic readout of a multipixel superconducting nanowire single photon detector}

\author{Johannes Tiedau,\authormark{1,*}, Timon Schapeler,\authormark{1} Vikas Anant,\authormark{2} Helmut Fedder,\authormark{3} Christine Silberhorn,\authormark{1} and Tim J. Bartley\authormark{1}}

\address{\authormark{1}Applied Physics, University of Paderborn, Warburger Stra\ss e 100, 33098 Paderborn, Germany\\
\authormark{2}Photon Spot, Inc. 142 W Olive Ave, Monrovia, CA 91016, USA\\
\authormark{3}Swabian Instruments GmbH,  Stammheimer Str. 41,70435 Stuttgart, Germany}

\email{\authormark{*}johannes.tiedau@uni-paderborn.de} 



\begin{abstract}
We present a time-over-threshold readout technique to count the number of activated pixels from an array of superconducting nanowire single photon detectors (SNSPDs). This technique maintains the intrinsic timing jitter of the individual pixels, places no additional heatload on the cryostat, and retains the intrinsic count rate of the time-tagger. We demonstrate proof-of-principle operation with respect to a four-pixel device. Furthermore, we show that, given some permissible error threshold, the number of pixels that can be reliably read out scales linearly with the intrinsic signal-to-noise ratio of the individual pixel response.
\end{abstract}

\section{Introduction}
Superconducting nanowire single photon detectors (SNSPDs) have become increasingly prevalent across many areas of optical sensing~\cite{natarajan_superconducting_2012}. They are particularly suited in applications which require high efficiency, low noise and low jitter when measuring low numbers of photons~\cite{marsili_detecting_2013,esmaeil_zadeh_single-photon_2017,korzh_demonstrating_2018}. These devices do not resolve the number of photons; rather, they fire if at least one photon is incident. Quasi-photon-number-resolution can be obtained by building arrays of these devices~\cite{dauler_multi-element_2007,divochiy_superconducting_2008,marsili_physics_2009,jahanmirinejad_photon-number_2012,rosenberg_high-speed_2013,zhao_superconducting-nanowire_2013,verma_four-pixel_2014,najafi_-chip_2015,chen_16-pixel_2018}; counting the number of detectors which fire places a lower bound on the number of photons which are incident. Moreover, placing these devices in spatial arrays means that the incoming light beam may be imaged~\cite{miki_64-pixel_2014,shaw_arrays_2015,allman_near-infrared_2015}. As the size of these arrays increases, a key challenge is to read out each  pixel, whilst maintaining the attractive properties of the constituent detectors. Since SNSPDs require cryogenic temperatures to operate, increasing the number of pixels typically increases the number of readout channels, which in turn increases the heatload of the cryogenic system. 

A variety of readout schemes is presented in~\cite{mccaughan_readout_2018}. A distinction is made between on-chip superconducting signal processing, which may be performed by {e.g.} single-flux quantum (SFQ) logic (see {e.g.}~\cite{yamashita_crosstalk-free_2012,hofherr_orthogonal_2012}) and analogue on chip multiplexing followed by off-chip processing, such as time~\cite{hofherr_time-tagged_2013,zhu2018scalable} and frequency~\cite{doerner_frequency-multiplexed_2017} signature read-out.


In the context of photon counting, knowing which pixel information is less important than knowing how many pixels fire. This can be achieved by connecting the output of each detector in series, such that the height of the output pulse is linearly proportional to the sum of the number of pixels which fire. This can be read out with a single electrical readout line, however analysis of the pulse requires a multilevel discriminator, which may increase the jitter, or splitting the readout line into several separate time-tagger channels, which quickly increases the hardware requirements.

To address this, in this letter we present a scalable single channel readout scheme which can extract the number of pixels firing in a multipixel array. This is achieved by measuring both the rising- and falling edge of the electrical pulse arising from a series connection of the outputs of each pixel, using a suitable time tagger. The arrival time information is retained in the rising edge, which maintains the intrinsic low jitter of the detectors. The falling edge is related to the recovery time of the detectors, which increases depending on the number of detectors which fire. Therefore, the time difference between the rising and falling edges, i.e. the pulse duration, is also proportional to the number of activated pixels. This implements a rudimentary time-over-threshold circuit, as has been employed in other contexts~\cite{fujiwara_new_2008,yonggang_linear_2014}. As such, information for both the number of pixels and the photons arrival time is preserved and read out on a single channel, in principle independent of the size of the array. 

\section{Measurement Scheme}
\begin{figure}[h] 
    \centering
    \includegraphics[width=\textwidth]{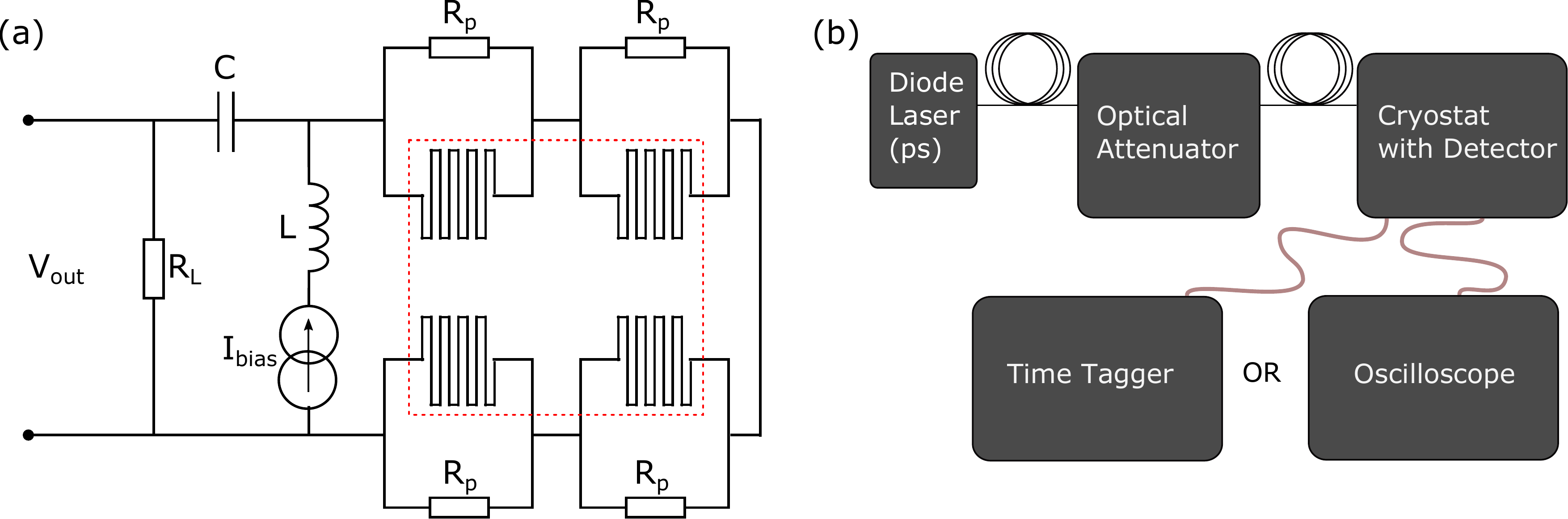}
    \caption{(a) Circuit diagram of a four-pixel SNSPD (marked in red). Each detector is connected in parallel with a resistor $R_p$ and connected to a constant current source $I_\text{bias}$ (via a bias-tee of inductance $L$ and capacitance $C$). The voltage $V_\text{out}$ across a load resistor $R_L$ is measured. All detector areas are connected in series to enable a single-channel read-out. (b) A fiber-coupled diode laser produces light at 1550~nm which is then attenuated and detected with a four-pixel SNSPD at 0.77~K. The electronic response from the detector is either measured with a timetagger that is connected with a single electric channel, or an oscilloscope.}
    \label{fig_scheme}
\end{figure}

To demonstrate this approach, we use a four-element multipixel detector wired in series  (Fig.~\ref{fig_scheme}~(a)) and connect it to the time-tagger. A schematic of the setup is shown in Fig.~\ref{fig_scheme}~(b). 
We generate coherent states with a fiber-coupled diode laser  (PicoQuant) with a repetition rate of 100~kHz centered around 1550~nm and a pulse duration of 50~ps. This light is  attenuated and detected with a four-pixel superconducting nanowire detector (PhotonSpot) which is able to generate electrical pulses with four different amplitudes depending on the number of pixels which detect photons.  The electronic response is either measured with a TimeTagger Ultra (Swabian Instruments) that is connected with a single electric channel, or an oscilloscope which is used as a reference. 

\begin{figure}[h]    \centering
    \includegraphics[width=\textwidth]{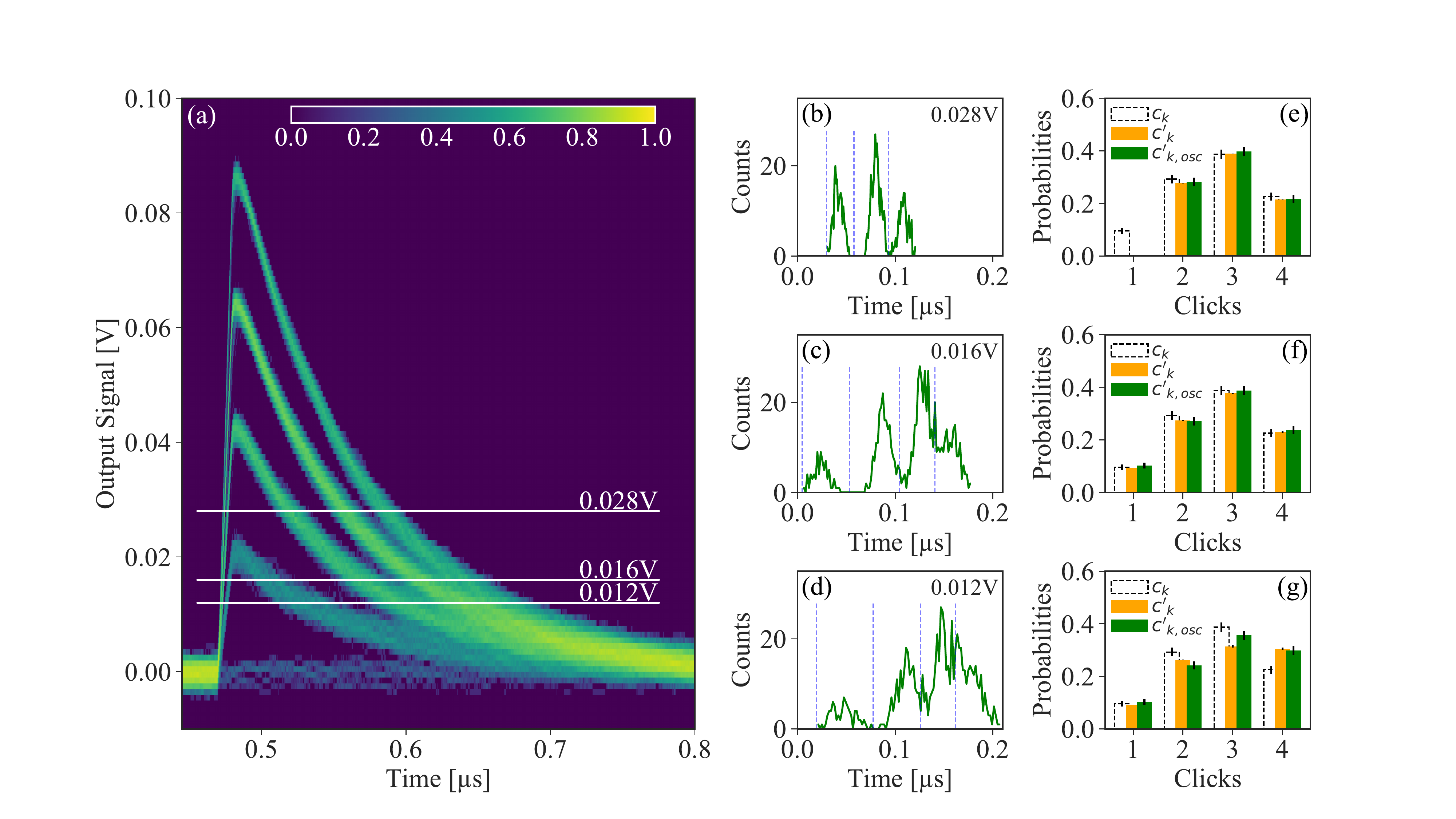}
    \caption{Oscilloscope traces showing the electric response of the four-pixel detector (a) as an intensity heat-map. As an example the evaluation of the output signal is shown for three different thresholds in subfigures (b-d). Here the time difference between the first rising slope and the first falling slope at the specific threshold is plotted as a histogram. Subfigures (e-g) shows the comparison between the click probabilities from the oscilloscope reference measurement ($c_k$, based on pulse height), the time over threshold value from the oscilloscope ($c'_{k,\text{OSC}}$) and the time over threshold value from the timetagger ($c'_k$). Compare text for further details.}
    \label{fig:osci}
\end{figure}
An example of the oscilloscope trace for such a device is shown in Fig.~\ref{fig:osci}~(a). If we neglect the risetime of the device, we can model this trace by treating each additional pixel as adding an exponential decay, modulated by some noise of width $\sigma_v$. That is, the detector response function for $n$ pixels firing scales as $n A e^{-\tau/t}$, where $A$ is the amplitude of a single pixel, and $\tau$ is a characteristic decay constant, independent of the number of pixels. The constants $A$, $\tau$ and $\sigma_v$ depend on the electronics of the detector circuitry. To evaluate the number of pixels which fire using a measurement in the time domain, we seek the time taken to decay to a particular threshold $a_0$. This occurs at 
\begin{equation}\label{eq:response}
t\left(n\right)=-\tau\ln \frac{a_0}{nA}
\end{equation}
Thus we have mapped the number of pixels firing to the time above threshold $a_0$ for this simple device. This equation also demonstrates the scaling of the maximum count rate for this device, namely logarithmically with the number of pixels $n$. 

Experimentally, this is achieved by the timetagger recording timestamps if the incoming electrical signal exceeds (rising edge) and deceeds (falling edge) a defined discrimination level. We investigated 25 different thresholds. Furthermore, in our implementation, an artificial deadtime of 400~ns is set on the timetagger to avoid extra triggering from electrical noise. The time-difference between a rising edge and a falling edge is calculated for all detection events and used as a measure for the pulse height.

\section{Click probability evaluation}
Based on the oscilloscope and time tagger measurement the click probability can be extracted in three different ways. For the oscilloscope traces it is straightforward to analyse the pulse height to determine the click probability $c_k$. Fig.~\ref{fig:osci}~(a) shows that the response functions for the four detection cases can be well separated at $0.5~\mu s$. While this method is capable of detecting the click probability with high accuracy, it requires sufficient voltage resolution as well as timing resolution. As many readout schemes involve timetaggers that can only record events at a specific threshold we investigate a readout scheme without any voltage resolution. As an example the time difference between the first rising edge and the first falling edge at a specific threshold is shown in Fig.~\ref{fig:osci}~(b-d) as a histogram. Counts in a specific area are integrated and ascribe the click probability $c'_{k\textrm{,osc}}$.

The time-over-threshold value can also be recorded directly with a timetagger, bypassing the oscilloscope entirely, and analysed in a similar way, to extract the click probability $c'_k$. The comparison between all three methods can be seen in Fig.~\ref{fig:osci}~(e-g) showing good consistency between the methods. A more detailed error analysis of the time-over-threshold scheme is given in the next section. The form of the measured click distribution is given by the Poissonian photon number distribution of a coherent state convolved with the four-element multiplexing detector response.


\section{Error Analysis}
In order to determine the quality of our new method we compare the time tagger measurement with the measured oscilloscope traces as shown in Fig.~\ref{fig:osci}~(a). 

\begin{figure}[H]
    \centering
    \includegraphics[width=\textwidth]{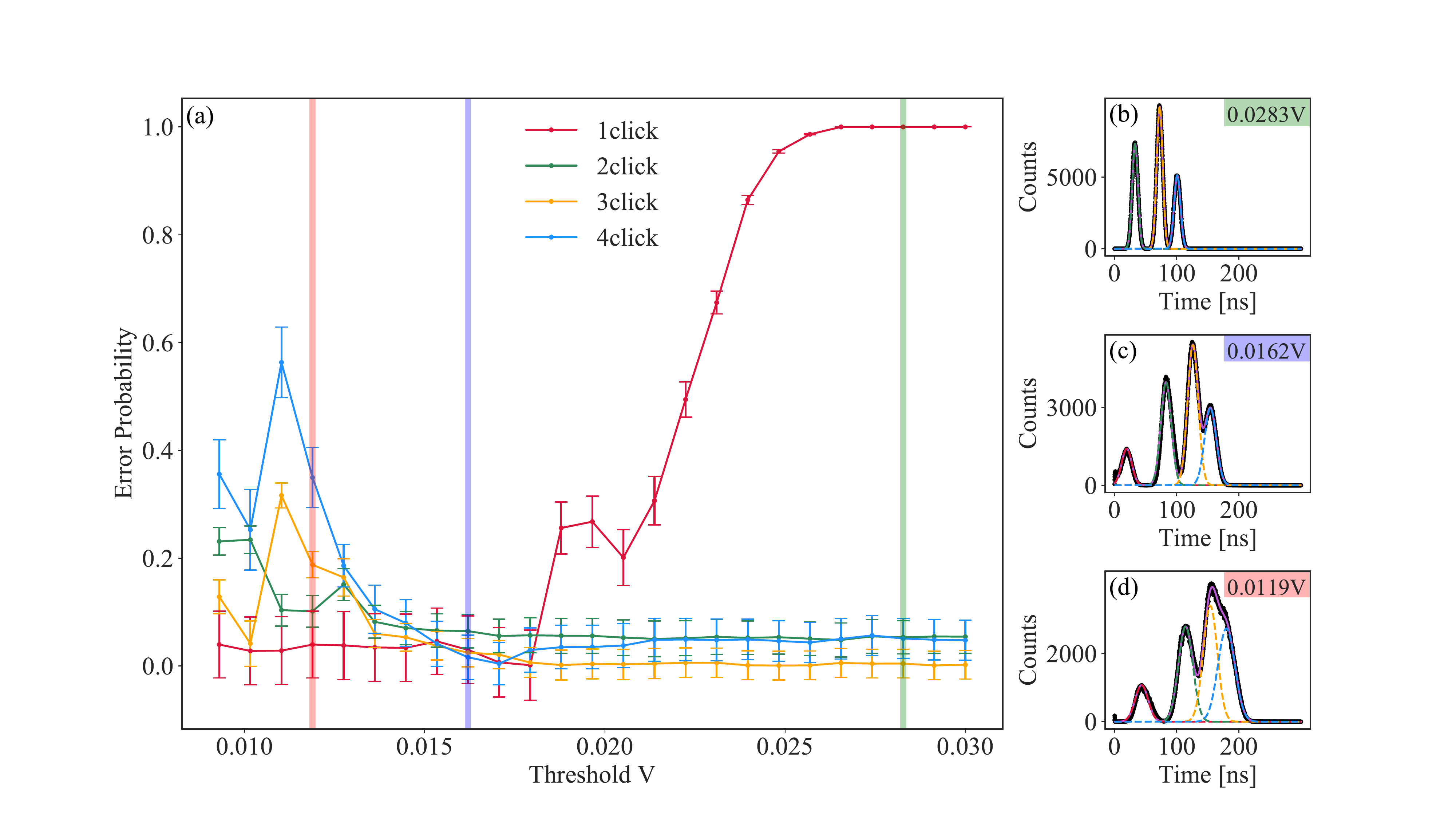}
    \caption{Comparison between click probabilities from oscilloscope traces and timetagger histograms for an ensemble measurement (a). (b)-(d) Start-stop histograms for three example thresholds are shown including Gaussian fits to extract the click probability $c'_k$. Compare text for further information.}
    \label{fig:errorprob}
\end{figure}

Fig.~\ref{fig:errorprob} shows the comparison between the oscilloscope reference measurement and the timetagger measurement. Start-stop histograms from the timetagger are analysed for 25 different trigger thresholds (three examples for 0.0283~V, 0.0162~V and 0.0119~V are shown in Fig.~\ref{fig:errorprob}~(b) - (d) respectively) and Gaussian functions are fitted to the histogram (colored-dashed lines in Fig.~\ref{fig:errorprob}~(b) - (d)). We will label the $k$th Gaussian fit $g_k$. These fits are then integrated to extract the click probability $c'_k$ for having $k$ simultaneous detection events. In order to compare these probabilities we calculated the error probability as the relative deviation of these two values
\begin{equation}
    p_{\text{error}, k} = \frac{\left|c_k - c'_k\right|}{c_k}
\end{equation}
which is shown in Fig.~\ref{fig:errorprob}~(a). It can be seen that at low thresholds, the error probability is quite high as the Gaussian functions are highly overlapping. For high thresholds on the other hand the two, three and four-fold events can be distinguished precisely but the threshold is above the maximum voltage for the single click event (compare oscilloscope trace in Fig.~\ref{fig:osci}). An optimal threshold can be found around 0.0162~V where the peaks of the two, three and four event clicks can be identified and the single click event is still fully visible. 

The presented approach based on Gaussian fits requires ensemble measurements. This is a valid approach for example for state characterisation where the same state can be investigated repeatedly. For some applications, however, single-shot experiments should be considered, in which the requirements may be more demanding. Here, the task is to estimate directly the  number of incident photons for each individual run of the experiment, for example when heralding single photons in the presence of multiphoton events. For this reason we introduced time windows in the timetagger histogram. If one start-stop event has a time duration within one of these windows we consider this as an event corresponding to this time window. The upper and lower limits of these time windows are given by the center positions between the Gaussian fit maxima. We will denote these limits with $u_k$ and $l_k$ for the upper and lower limit of the $k$th time window respectively. 

In contrast to ensemble measurements, two sources of error can appear for the single-shot case. As an example we consider the second time window which should identify events where two pixels have detected a photon simultaneously. Firstly, we can miss a two-pixel event if the time duration was too short (and therefore was identified as a one-pixel event) or too long (and therefore was identified as a three-click event). Secondly, we can misidentify a one-,three or four-click event as a two-click event if the start-stop time has the appropriate length.
We can use the Gaussian fits $g_k$ from ensemble measurements to estimate the magnitude of these two errors. In general we can write
\begin{equation}
\label{eq:missing}
    p_{\text{missing click, } k} =1- \frac{1}{M'_k}\int_{l_k}^{u_k} g_k
\end{equation}
and 
\begin{equation}\label{eq:misident}
    p_{\text{misidentified click, } k} = \frac{1}{N'_k}\sum_{i\neq k}\int_{l_k}^{u_k} g_{i}
\end{equation}
with normalization constants $M'_k$ and $N'_k$. If the trigger threshold from the timetagger is low enough such that all events are detected and all Gaussian fits are well separated, these normalisation factors would be equivalent $M'_k = N'_k$. However, physical imperfections causing overlapping Gaussian fits and thresholds above the maximal voltage of an event require careful normalisation as shown in Appendix~\ref{sec:normalization}. 

\begin{figure}[H]
    \centering
    \includegraphics[width=\textwidth]{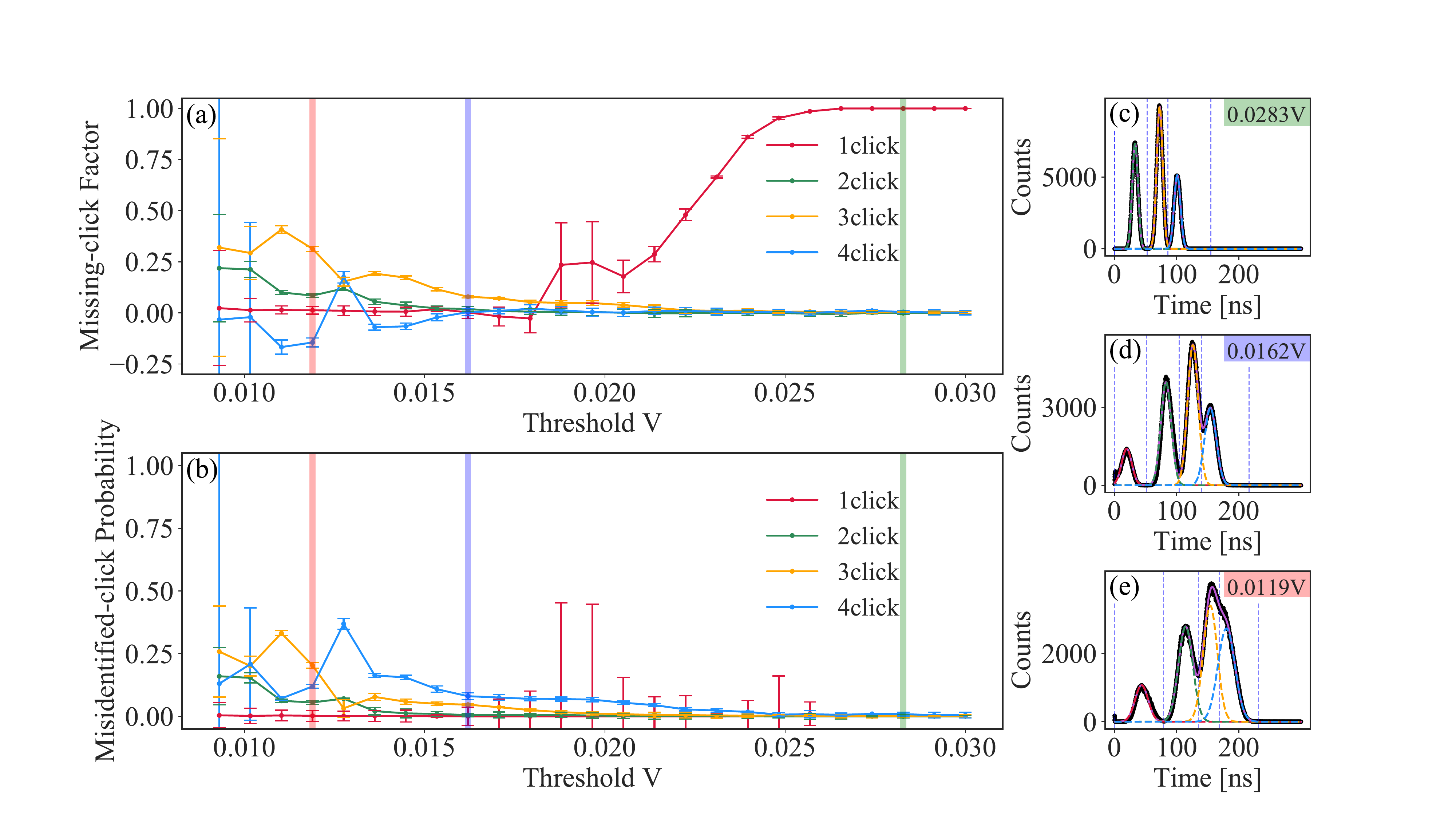}
    \caption{We consider two potential errors for single-shot measurements. (a) Probability of a missing a click and (b) probability of a misidentified click. Compare text for further details. Subfigures (b)-(d) show start-stop histograms for three example thresholds including the time windows (marked with horizontal lines).}
    \label{fig:errors}
\end{figure}

Both errors are shown in Fig.~\ref{fig:errors}. For the single-click event it can be seen that lower thresholds are well suited to detect these events. The misidentified click probability is low due to the large separation of the single- and two-click events in the histogram. For high thresholds it can be seen that an increasing number of clicks are missed as the threshold surpasses the maximum voltage for this event. At very high thresholds (not shown) the error probabilities for the two-,three- and four-event cases would also increase. Negative values for the missing click factor indicate that the calculated probability is higher than the corresponding normalisation constant $M'_k$.
This effect is very pronounced for the four-fold event at low thresholds. Here the the overlap of three-fold and four-fold events causes an overestimation of the four-fold rate. 
For the misidentified click probability it is beneficial to choose high thresholds. Especially the four-click event profits from higher thresholds as three- and four-fold events have the highest overlap. Misidentified click probabilities above $>0.0257$~V cannot be plotted anymore for the single event case because the normalisation diverges (no single clicks can be identified). 
Again a threshold around 0.0171~V seems to be optimal to reduce both errors simultaneously. Details about the calculations of the error bars can be found in the Appendix~\ref{sec:errors}.


\section{Scaling}
It is prudent to consider the scaling of this readout scheme as the number of individual pixels increases. 
 From Eq.~\ref{eq:response}, the behaviour of the response function is such that the time gaps between the response of each pixel get smaller the more pixels fire, {i.e.} the ratio $t\left(n+1\right)/t\left(n\right)$ decreases with increasing $n$. Therefore, for a fixed uncertainty in the time $t$, it becomes harder to distinguish higher numbers of pixels.

In general, the uncertainty in the measurement time $\sigma_t$ is the quadrature sum of the contributions from the detector itself and jitter of the time tagger, however we neglect the jitter as it is typically much smaller than the noise arising from the detector recovery slope. Using the exponential decay model of the response given by Eq.~\ref{eq:response}, we therefore seek the noise in the time domain $\sigma_t$ caused by noise in the voltage response $\sigma_v$. This is the product of $\sigma_v$ and the absolute value of the time derivative of the response function evaluated at a time $t$, {i.e.}
\begin{equation}\label{eq:sigmat}
   \sigma_t=\sigma_v \left|\frac{\mathrm{d}}{\mathrm{d}t}n A e^{-\tau/t}\bigg|_{t=t\left(n\right)}\right|=\frac{\sigma_va_0}{\tau}~,
\end{equation}
which is independent of bin number $n$. We consider the noise on the readout to be follow a Gaussian distribution of width $\sigma_t$ and centre positions for each $n$ pixels given by equation~\ref{eq:response}.

To evaluate the scalability of this readout scheme in principle, we seek the largest number of pixels $n_\textrm{max}$ that can be successfully read out in this manner, depending on the detector parameters noise $\sigma_v$, pixel height $A$, and decay time $\tau$. Our figure of merit is the misidentified click probability (Eq.~\ref{eq:misident}) of the $n_\textrm{max}-1$ bin, which is the ``worst case'', since we consider error contributions from $n_\textrm{max}$, as well as $n_\textrm{max}-2$. We further assume equal likelihood of each bin $n$ being occupied. A more thorough analysis should consider all error terms from $n_\textrm{max}-3...1$, however we choose to define $n_\textrm{max}$ such that the overlaps from $n_\textrm{max}-3$ (and therefore all smaller bins) is negligible. Furthermore, one may also wish to take into account the probabilities for the number of pixels firing, which is given by the convolution of the photon statistics of the incident beam and the splitting function of the multipixel detector (see {e.g.}~\cite{achilles_photon-number-resolving_2004,sperling_true_2012}).

To evaluate Eq.~\ref{eq:misident} explicitly, we begin with the Gaussian distribution functions $g_i$, with $i\in \left(n_\text{max},n_\text{max}-1,n_\text{max}-2\right)$. For a mean $\mu=-\tau\ln \frac{a_0}{iA}$ and standard deviation $\sigma=\frac{\sigma_v\tau}{a_0}$, this is given by
\begin{equation}
    g_i=\frac{a_0}{\sqrt{2\pi}\tau\sigma_v}e^{-\frac{a_0^2}{2\tau^2\sigma_v^2}\left(t+\tau\text{ln}\left[\frac{a_0}{i A}\right]\right)^2}~.
\end{equation}
We consider the optimal threshold value $a_0=A-\sigma_v$, since this provides the highest threshold whilst remaining below the noise of the first peak $A$. We further define the signal-to-noise ratio $\beta=A/\sigma_v$ as a key detector parameter, such that $a_0=\sigma_v\left(\beta-1\right)$. Making this substitution and applying Eq.~\ref{eq:misident}, we find that
\begin{equation}
    p_{\textrm{misidentified click},n_{\text{max}-1}}=1-\frac{\text{Erf}\left[\frac{(\beta -1) }{2 \sqrt{2}}\text{ln}\left[\frac{n_\text{max}}{n_\text{max}-1}\right]\right]+\text{Erf}\left[\frac{(\beta-1 ) }{2 \sqrt{2}}\text{ln}\left[\frac{n_\text{max}-1}{n_\text{max}-2}\right]\right]}{\text{Erf}\left[\frac{(\beta -1) }{2 \sqrt{2}}\text{ln}\left[\frac{n_\text{max}^2}{(n_\text{max}-1)(n_\text{max}-2)}\right]\right]+\text{Erf}\left[\frac{(\beta -1) }{2 \sqrt{2}}\text{ln}\left[\frac{n_\text{max}(n_\text{max}-1)}{(n_\text{max}-2)^2}\right]\right]}~,
\end{equation}
which is entirely dependent on the signal-to-noise parameter $\beta$ and $n_\text{max}$, and independent of the decay time $\tau$.  Indeed, this function is closely approximated by the Gaussian distribution given by
\begin{equation}
     p_{\textrm{misidentified click},n_{\text{max}-1}}\approx\frac{2}{3}e^{-\frac{\left(\beta-1\right)^2}{2\varsigma^2}}~,
\end{equation}
where the width $\varsigma=a+bn_\text{max}$, and $a=-1.69$ and $b=1.64$ are numerically determined constants. Thus a direct relationship can be found between the detector parameter $\beta$ and the maximum number of pixels that can be distinguished, up to a given error probability. Furthermore, we note that additional electronics could be implemented to change the detector response function and thus the noise scaling. For example, a linear (rather than exponential) decay would lead to an error probability independent of $n_\textrm{max}$.

\begin{figure}[H]
    \centering
    \includegraphics[width=0.6\textwidth]{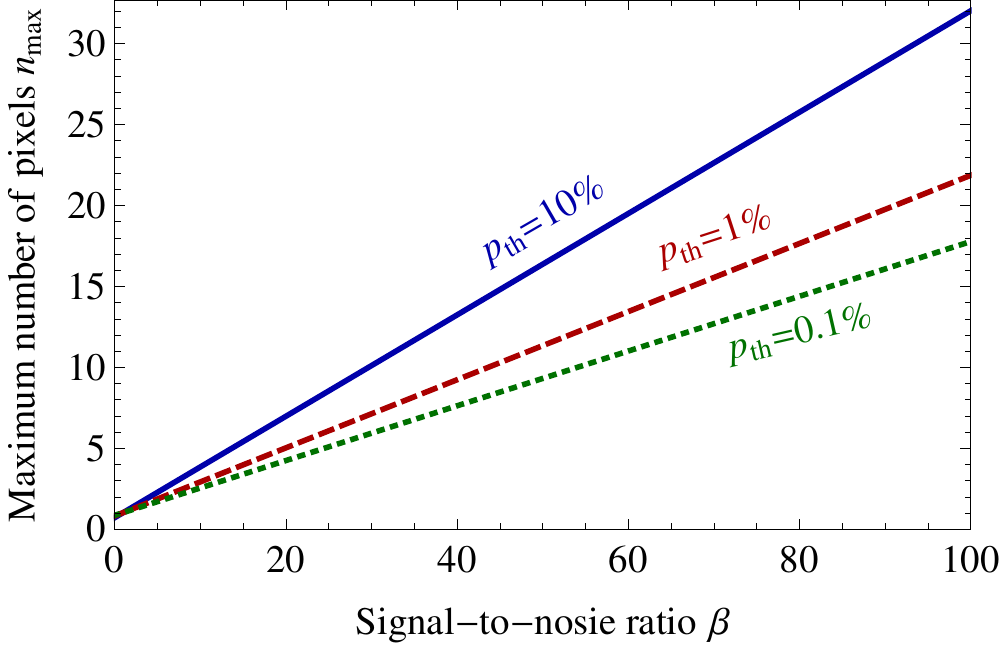}
    \caption{Maximum pixel number as a function of the signal-to-noise ratio parameter $\beta$. Each line indicates the detector quality that must be achieved in order to reach a particular error threshold.}
    \label{fig:scaling}
\end{figure}

If we tolerate errors of up to a given threshold $p_\text{th}$ when identifying pixels in a single-shot configuration, the maximum number of pixels that can be distinguished, as a function of the signal to noise of the detector $\beta$ is very closely approximated by
\begin{equation}
    n_\text{max}\approx\frac{1}{b}\left(\frac{\beta-1}{\sqrt{2\text{ln}\left[\frac{2}{3p_\text{th}}\right]}}-a\right)~,
\end{equation}
with the constants $a$ and $b$ as determined above. Thus the scaling of this readout scheme depends linearly on the signal to noise ratio of the detector, as illustrated in Fig.~\ref{fig:scaling}. This indicates that any measurement technique which increase the signal-to-noise ratio, such as the use of a box-car amplifier, will improve the overall scaling of the device.

Based on the signal to noise ratio of 10 in our experiment, we were able to distinguish 4 pixels with an error rate around 10\%, which is in good agreement with the analysis above.

\section{Conclusion}
We have presented a proof-of-principle demonstration of a scheme to read out a multipixel detector with a single electrical line, based on the time-over-threshold pulse length of the device when each pixel is connected in series. Based on a simple model of the response function of the device and its noise characteristics, we showed that the scaling of the device is depends linearly on the signal-to-noise ratio of readout. Therefore, further improvements to the intrinsic circuitry or amplification, which increase the signal to noise level, are particularly useful at higher pixel numbers. 

With this method, we have shown that the total number of fired pixels can be counted, based on the same response from each pixel. However, one could consider combining this technique with a pixel-dependent voltage response, such as shown by Gaggaero et al~\cite{gaggero2019amplitude}, to enable the extraction of which-pixel information, which is required for imaging applications.

\section*{Appendix}
\setcounter{section}{0}
\def\thesection{\Alph{section}}

\section{Normalization Factors}
\label{sec:normalization}
The missing click factor $ p_{\text{missing click, } k}$ should indicate that events are missed, which may be caused by a threshold level $a_0$ above the detector output voltage $A$. In order to capture this behavior the normalisation constant $M'_k$ should be calculated for a fixed threshold $a_0$ that is lower than the corresponding maximal voltage $k\cdot A$ for this $k$-fold event. We suggest to calculate the normalisation constant $M'_k$ for a threshold $a_0$ between the voltage level of the $k-1$ and $k$th level
\begin{equation}
    M'_k = \int g_k\Big| _{a_0 = kA-\frac{A}{2}}~.
\end{equation}
For our calculations all threshold for $M'_{k\neq1}$ were calculated at $a_0 = 2A-\frac{A}{2}$ as the error probability for this threshold is already low enough (compare Fig.~\ref{fig:errorprob}). 


The normalisation for the misidentified click probability $p_\text{misidentified click, k}$ is different as we here consider a post-selected click probability. We are interested in the misidentified click probability inside the subset of having $k$ events. The size of this subset, which is equivalent to the desired normalisation constant $N'_k$, is given by the total number of events recognised in the $k$th time window \begin{equation}
N'_k= \sum_{k}\int_{l_k}^{u_k} g_{k}~.
\end{equation}

\section{Measurement Uncertainties}
\label{sec:errors}
The error bars shown in Fig.~\ref{fig:errorprob} and  Fig.~\ref{fig:errors} are based on Gaussian error propagation resulting in
\begin{align}
    \sigma_{p_{\text{error},k}} &= \sqrt{\left(\frac{\partial p_{\text{error},k}}{\partial c_k}\cdot\sigma_{c_k}\right)^2 + \left(\frac{\partial p_{\text{error},k}}{\partial c'_k}\cdot\sigma_{c'_k}\right)^2} \\
    \sigma_{p_{\text{missing},k}} &= \sqrt{\left(\frac{\partial p_{\text{missing},k}}{\partial M'_k}\cdot\sigma_{M'_k}\right)^2 + \left(\frac{\partial p_{\text{missing},k}}{\partial g_k}\cdot\sigma_{g_k}\right)^2}\\
    \sigma_{p_{\text{misidentified},k}} &= \sqrt{\left(\frac{\partial p_{\text{misidentified},k}}{\partial N'_k}\cdot\sigma_{N'_k}\right)^2 + \left(\frac{\partial p_{\text{misidentified},k}}{\partial g_k}\cdot\sigma_{g_k}\right)^2}~.
\end{align}
Here the error $\sigma_{c_k}$ is based on the statistical uncertainties of the measurement whereas $\sigma_{c'_k}$ is based on the fit uncertainty. The experimentally measured histogram gives a discrete function $h_{\text{exp}}$ which is fitted with Gaussian functions $g_k$. The fitting uncertainties are used to calculate a maximal value for the fit denoted as $g^\text{max}_k$. We consider the fitting uncertainty and the distance of the experimental data to the histogram for the overall uncertainty of $g_k$ given by:
\begin{equation}
    \sigma_{g_{k}}^2 = \left| \int g_k - \int g^\text{max}_k \right| + \left| \sum h_\text{exp} -  \int g_k \right|
\end{equation}
If the fits are only evaluated in a specific region as used in Eq.~\ref{eq:missing} then the overall uncertainty of $g_k$ is only evaluated in this region. E.g.
\begin{equation}
    \sigma_{g_{k}}^2 = \left| \int^{u_k}_{l_k} g_k - \int^{u_k}_{l_k} g^\text{max}_k \right| + \left| \sum^{u_k}_{l_k} h_\text{exp} -  \int^{u_k}_{l_k} g_k \right|
\end{equation}

\section*{Funding}\par

This research is supported by the EU H2020-FETFLAG-2018-03 under Grant Agreement No. 820365 (PhoG) and the German Federal Ministry of Education and Research (BMBF) under the ``Quantum Futur'' Programme, project number 13N14911.

\section*{Disclosures}\par
JT, CS, TJB: (P), VA: Photon Spot (I,E,P), HF: Swabian Instruments GmbH (I,E,P) 



\bibliography{Single-channel_readout.bib}

\begin{thebibliography}{10}
\newcommand{\enquote}[1]{``#1''}

\bibitem{natarajan_superconducting_2012}
C.~M. Natarajan, M.~G. Tanner, and R.~H. Hadfield, \enquote{Superconducting
  nanowire single-photon detectors: physics and applications,}
  {\protect\JournalTitle{Superconductor Science and Technology}} \textbf{25},
  063001 (2012).

\bibitem{marsili_detecting_2013}
F.~Marsili, V.~B. Verma, J.~A. Stern, S.~Harrington, A.~E. Lita, T.~Gerrits,
  I.~Vayshenker, B.~Baek, M.~D. Shaw, R.~P. Mirin, and S.~W. Nam,
  \enquote{Detecting single infrared photons with 93\% system efficiency,}
  {\protect\JournalTitle{Nature Photonics}} \textbf{7}, 210--214 (2013).

\bibitem{esmaeil_zadeh_single-photon_2017}
I.~Esmaeil~Zadeh, J.~W.~N. Los, R.~B.~M. Gourgues, V.~Steinmetz, G.~Bulgarini,
  S.~M. Dobrovolskiy, V.~Zwiller, and S.~N. Dorenbos, \enquote{Single-photon
  detectors combining high efficiency, high detection rates, and ultra-high
  timing resolution,} {\protect\JournalTitle{APL Photonics}} \textbf{2}, 111301
  (2017).

\bibitem{korzh_demonstrating_2018}
B.~A. Korzh, Q.-Y. Zhao, S.~Frasca, J.~P. Allmaras, T.~M. Autry, E.~A. Bersin,
  M.~Colangelo, G.~M. Crouch, A.~E. Dane, T.~Gerrits, F.~Marsili, G.~Moody,
  E.~Ramirez, J.~D. Rezac, M.~J. Stevens, E.~E. Wollman, D.~Zhu, P.~D. Hale,
  K.~L. Silverman, R.~P. Mirin, S.~W. Nam, M.~D. Shaw, and K.~K. Berggren,
  \enquote{Demonstrating sub-3 ps temporal resolution in a superconducting
  nanowire single-photon detector,} {\protect\JournalTitle{arXiv preprint
  arXiv:1804.06839}}  (2018).

\bibitem{dauler_multi-element_2007}
E.~A. Dauler, B.~S. Robinson, A.~J. Kerman, J.~K.~W. Yang, K.~M. Rosfjord,
  V.~Anant, B.~Voronov, G.~Gol'tsman, and K.~K. Berggren,
  \enquote{Multi-{Element} {Superconducting} {Nanowire} {Single}-{Photon}
  {Detector},} {\protect\JournalTitle{IEEE Transactions on Applied
  Superconductivity}} \textbf{17}, 279--284 (2007).

\bibitem{divochiy_superconducting_2008}
A.~Divochiy, F.~Marsili, D.~Bitauld, A.~Gaggero, R.~Leoni, F.~Mattioli,
  A.~Korneev, V.~Seleznev, N.~Kaurova, O.~Minaeva, G.~Gol'tsman, K.~G.
  Lagoudakis, M.~Benkhaoul, F.~Lévy, and A.~Fiore, \enquote{Superconducting
  nanowire photon-number-resolving detector at telecommunication wavelengths,}
  {\protect\JournalTitle{Nature Photonics}} \textbf{2}, 302--306 (2008).

\bibitem{marsili_physics_2009}
F.~Marsili, D.~Bitauld, A.~Gaggero, S.~Jahanmirinejad, R.~Leoni, F.~Mattioli,
  and A.~Fiore, \enquote{Physics and application of photon number resolving
  detectors based on superconducting parallel nanowires,}
  {\protect\JournalTitle{New Journal of Physics}} \textbf{11}, 045022 (2009).

\bibitem{jahanmirinejad_photon-number_2012}
S.~Jahanmirinejad, G.~Frucci, F.~Mattioli, D.~Sahin, A.~Gaggero, R.~Leoni, and
  A.~Fiore, \enquote{Photon-number resolving detector based on a series array
  of superconducting nanowires,} {\protect\JournalTitle{Applied Physics
  Letters}} \textbf{101}, 072602 (2012).

\bibitem{rosenberg_high-speed_2013}
D.~Rosenberg, A.~J. Kerman, R.~J. Molnar, and E.~A. Dauler, \enquote{High-speed
  and high-efficiency superconducting nanowire single photon detector array,}
  {\protect\JournalTitle{Optics Express}} \textbf{21}, 1440--1447 (2013).

\bibitem{zhao_superconducting-nanowire_2013}
Q.~è. Zhao, A.~McCaughan, F.~Bellei, F.~Najafi, D.~De~Fazio, A.~Dane, Y.~Ivry,
  and K.~K. Berggren, \enquote{Superconducting-nanowire single-photon-detector
  linear array,} {\protect\JournalTitle{Applied Physics Letters}} \textbf{103},
  142602 (2013).

\bibitem{verma_four-pixel_2014}
V.~B. Verma, R.~Horansky, F.~Marsili, J.~A. Stern, M.~D. Shaw, A.~E. Lita,
  R.~P. Mirin, and S.~W. Nam, \enquote{A four-pixel single-photon
  pulse-position array fabricated from {WSi} superconducting nanowire
  single-photon detectors,} {\protect\JournalTitle{Applied Physics Letters}}
  \textbf{104}, 051115 (2014).

\bibitem{najafi_-chip_2015}
F.~Najafi, J.~Mower, N.~C. Harris, F.~Bellei, A.~Dane, C.~Lee, X.~Hu,
  P.~Kharel, F.~Marsili, S.~Assefa, K.~K. Berggren, and D.~Englund,
  \enquote{On-chip detection of non-classical light by scalable integration of
  single-photon detectors,} {\protect\JournalTitle{Nature Communications}}
  \textbf{6}, 5873 (2015).

\bibitem{chen_16-pixel_2018}
Q.~Chen, B.~Zhang, L.~Zhang, R.~Ge, R.~Xu, Y.~Wu, X.~Tu, X.~Jia, L.~Kang,
  J.~Chen, and P.~Wu, \enquote{A 16-pixel {NbN} nanowire single photon detector
  coupled with 300 micrometer fiber,} {\protect\JournalTitle{arXiv preprint
  arXiv:1811.09779}}  (2018).

\bibitem{miki_64-pixel_2014}
S.~Miki, T.~Yamashita, Z.~Wang, and H.~Terai, \enquote{A 64-pixel {NbTiN}
  superconducting nanowire single-photon detector array for spatially resolved
  photon detection,} {\protect\JournalTitle{Optics Express}} \textbf{22},
  7811--7820 (2014).

\bibitem{shaw_arrays_2015}
M.~D. Shaw, F.~Marsili, A.~D. Beyer, J.~A. Stern, G.~V. Resta, P.~Ravindran,
  S.~Chang, J.~Bardin, D.~S. Russell, J.~W. Gin, F.~D. Patawaran, V.~B. Verma,
  R.~P. Mirin, S.~W. Nam, and W.~H. Farr, \enquote{Arrays of {WSi}
  {Superconducting} {Nanowire} {Single} {Photon} {Detectors} for {Deep}-{Space}
  {Optical} {Communications},} in \emph{{CLEO}: 2015 (2015), paper {JTh}2A.68,}
   (Optical Society of America, 2015), p. JTh2A.68.

\bibitem{allman_near-infrared_2015}
M.~S. Allman, V.~B. Verma, M.~Stevens, T.~Gerrits, R.~D. Horansky, A.~E. Lita,
  F.~Marsili, A.~Beyer, M.~D. Shaw, D.~Kumor, R.~Mirin, and S.~W. Nam,
  \enquote{A near-infrared 64-pixel superconducting nanowire single photon
  detector array with integrated multiplexed readout,}
  {\protect\JournalTitle{Applied Physics Letters}} \textbf{106}, 192601 (2015).

\bibitem{mccaughan_readout_2018}
A.~N. McCaughan, \enquote{Readout architectures for superconducting nanowire
  single photon detectors,} {\protect\JournalTitle{Superconductor science \&
  technology}} \textbf{31} (2018).

\bibitem{yamashita_crosstalk-free_2012}
T.~Yamashita, S.~Miki, H.~Terai, K.~Makise, and Z.~Wang,
  \enquote{Crosstalk-free operation of multielement superconducting nanowire
  single-photon detector array integrated with single-flux-quantum circuit in a
  0.1\&\#{xA}0;{W} {Gifford}\&\#x2013;{McMahon} cryocooler,}
  {\protect\JournalTitle{Optics Letters}} \textbf{37}, 2982--2984 (2012).

\bibitem{hofherr_orthogonal_2012}
M.~Hofherr, O.~Wetzstein, S.~Engert, T.~Ortlepp, B.~Berg, K.~Ilin, D.~Henrich,
  R.~Stolz, H.~Toepfer, H.-G. Meyer, and M.~Siegel, \enquote{Orthogonal
  sequencing multiplexer for superconducting nanowire single-photon detectors
  with {RSFQ} electronics readout circuit,} {\protect\JournalTitle{Optics
  Express}} \textbf{20}, 28683--28697 (2012).

\bibitem{hofherr_time-tagged_2013}
M.~Hofherr, M.~Arndt, K.~Il'in, D.~Henrich, M.~Siegel, J.~Toussaint, T.~May,
  and H.~Meyer, \enquote{Time-{Tagged} {Multiplexing} of {Serially} {Biased}
  {Superconducting} {Nanowire} {Single}-{Photon} {Detectors},}
  {\protect\JournalTitle{IEEE Transactions on Applied Superconductivity}}
  \textbf{23}, 2501205--2501205 (2013).

\bibitem{zhu2018scalable}
D.~Zhu, Q.-Y. Zhao, H.~Choi, T.-J. Lu, A.~E. Dane, D.~Englund, and K.~K.
  Berggren, \enquote{A scalable multi-photon coincidence detector based on
  superconducting nanowires,} {\protect\JournalTitle{Nature nanotechnology}}
  \textbf{13}, 596 (2018).

\bibitem{doerner_frequency-multiplexed_2017}
S.~Doerner, A.~Kuzmin, S.~Wuensch, I.~Charaev, F.~Boes, T.~Zwick, and
  M.~Siegel, \enquote{Frequency-multiplexed bias and readout of a 16-pixel
  superconducting nanowire single-photon detector array,}
  {\protect\JournalTitle{Applied Physics Letters}} \textbf{111}, 032603 (2017).

\bibitem{fujiwara_new_2008}
T.~Fujiwara and H.~Takahashi, \enquote{A new multi-level time over threshold
  method for energy resolving multi-channel systems,} in \emph{2008 {IEEE}
  {Nuclear} {Science} {Symposium} {Conference} {Record},}  (2008), pp.
  3413--3415.

\bibitem{yonggang_linear_2014}
W.~Yonggang, C.~Xinyi, L.~Deng, Z.~Wensong, and L.~Chong, \enquote{A {Linear}
  {Time}-{Over}-{Threshold} {Digitizing} {Scheme} and {Its} 64-channel {DAQ}
  {Prototype} {Design} on {FPGA} for a {Continuous} {Crystal} {PET}
  {Detector},} {\protect\JournalTitle{IEEE Transactions on Nuclear Science}}
  \textbf{61}, 99--106 (2014).

\bibitem{achilles_photon-number-resolving_2004}
D.~Achilles, C.~Silberhorn, C.~Sliwa, K.~Banaszek, I.~A. Walmsley, M.~J. Fitch,
  B.~C. Jacobs, T.~B. Pittman, and J.~D. Franson,
  \enquote{Photon-number-resolving detection using time-multiplexing,}
  {\protect\JournalTitle{Journal of Modern Optics}} \textbf{51}, 1499 (2004).

\bibitem{sperling_true_2012}
J.~Sperling, W.~Vogel, and G.~S. Agarwal, \enquote{True photocounting
  statistics of multiple on-off detectors,} {\protect\JournalTitle{Physical
  Review A}} \textbf{85}, 023820 (2012).

\bibitem{gaggero2019amplitude}
A.~Gaggero, F.~Martini, F.~Mattioli, F.~Chiarello, R.~Cernansky, A.~Politi, and
  R.~Leoni, \enquote{Amplitude-multiplexed readout of single photon detectors
  based on superconducting nanowires,} {\protect\JournalTitle{Optica}}
  \textbf{6}, 823--828 (2019).

\end{thebibliography}






\end{document}